\documentclass[10pt,a4paper,twocolumn]{article}
\usepackage[margin=17mm,columnsep=6mm]{geometry}
\usepackage{times,amsmath,amssymb,bm,graphicx,booktabs,array,url}
\usepackage[numbers,sort&compress]{natbib}
\usepackage[hidelinks]{hyperref}
\makeatletter
\renewcommand\section{\@startsection{section}{1}{\z@}{-3.5ex\@plus -1ex\@minus -.2ex}{2.3ex\@plus .2ex}{\normalfont\large\bfseries\raggedright}}
\renewcommand\subsection{\@startsection{subsection}{2}{\z@}{-3.0ex\@plus -1ex\@minus -.2ex}{1.5ex\@plus .2ex}{\normalfont\normalsize\bfseries\raggedright}}
\makeatother
\newcommand{\pendulumMinimumTime}{0.952}
\newcommand{\pendulumMinimumDepth}{-0.74}

\begin{document}
\twocolumn[{
\begin{center}
{\Large\bfseries Resonant-pair energy return beyond the early Green--Kubo plateau in a quartic chain\par}
\vspace{10pt}
Julian Olszewski, Zijian Liu, Jie Xu\textsuperscript{*}\par
\vspace{5pt}
{\small Department of Mechanical and Industrial Engineering,\\
University of Illinois Chicago, 842 West Taylor Street (MC 251),\\
Chicago, Illinois 60607, USA\par}
\vspace{5pt}
{\small\textsuperscript{*}Corresponding author: \href{mailto:jiexu@uic.edu}{jiexu@uic.edu}\par}
{\small\textit{Author note.} Julian Olszewski and Zijian Liu contributed equally to this work.\par}
\end{center}
\vspace{3pt}
\begin{minipage}{\textwidth}
\noindent\textbf{Abstract.} The Green--Kubo formula gives the thermal
conductivity as the long-time limit of the running Green--Kubo integral,
the time integral of the equilibrium heat-current autocorrelation
function. In simulations, this integral often approaches a plateau at
finite time. We study how it evolves after such an early plateau in a
weakly coupled chain of quartic oscillators, where the plateau value is
close to the rate at which independent neighboring cells exchange energy.
Two neighboring cells with nearly equal frequencies form a resonant pair:
their relative phase oscillates like a pendulum, and energy that has
crossed the bond between them flows back. We show that, at weak coupling,
the running integral of such a pair follows the pendulum trapping response
of nonlinear Landau damping. For the quartic potential, the
small-oscillation period of this pendulum,
$T_{\rm lib}\propto\varepsilon^{-1/2}$ with intercell coupling
$\varepsilon$, does not depend on energy, so all resonant pairs share one
time scale. Without fitted parameters, the theory predicts that the running
integral of an isolated pair falls to a negative first minimum of
$\pendulumMinimumDepth$ times the plateau value at
$\pendulumMinimumTime\,T_{\rm lib}$. Simulations of isolated pairs approach
this prediction as the coupling decreases. In chains, the first minimum
occurs on the same time scale but is shallower, and a small positive value
can remain much later. Chains of strongly chaotic cells show no comparable
decrease, and energy-conserving noise much faster than $1/T_{\rm lib}$
largely removes the decrease. The pair response provides a quantitative
finite-time reference for interpreting early Green--Kubo plateaus in such
chains.
\par\medskip
\noindent\textbf{Running title:} Resonant-pair energy return in a quartic chain
\par\smallskip
\noindent\textbf{Keywords:} coupled oscillators; Green--Kubo relation; heat transport; nonlinear resonance; quartic chain
\end{minipage}
\vspace{14pt}
}]
\thispagestyle{plain}

\section{Introduction}
\label{sec:intro}

Thermal insulation and thermoelectric energy conversion motivate reliable
microscopic descriptions of heat transport in anharmonic solids~\cite{Knoop}.
A basic question is under what conditions microscopic dynamics produce
normal, diffusive heat transport~\cite{Dhar}. The question is especially
subtle in weakly coupled systems, where local energy exchange and
collective redistribution can occur on widely separated time scales. This
separation also matters for simulations with finite observation times,
whose transport estimates must resolve the relevant dynamical
scales~\cite{Knoop}.

The Green--Kubo formula gives the conductivity of an infinite system as the
long-time limit of the running Green--Kubo integral, the time integral of the equilibrium
heat-current autocorrelation function up to a cutoff time~\cite{Dhar}. In
simulations, this integral often approaches a plateau at finite
time~\cite{Knoop}. Such a plateau need not be the long-time limit, since
slower dynamics can later move the integral away from it. Related
distinctions between intermediate and asymptotic transport have been
reported for momentum-conserving chains near integrable limits~\cite{LLP20}.
In the weakly coupled anharmonic chain studied here, the early plateau is
close to the bare coefficient, the independent-cell term in the
weak-coupling coefficient of Bernardin et al.~\cite{BHLLO}. It is the rate
at which two neighboring cells exchange energy when each evolves as if it
were uncoupled. Descriptions of heat transport
by local oscillators also underlie minimum-conductivity models for
disordered solids~\cite{CP89,CWP92}. A quantitative interpretation of the early
plateau therefore requires the dynamics that change the integral after it,
and their time scales.

Rigorous weak-coupling results depend on the dynamics of the individual
cells. Dolgopyat and Liverani~\cite{DL} proved an energy-diffusion limit for
finitely many weakly coupled geodesic flows on manifolds of negative
curvature, and Liverani and Olla~\cite{LO}
obtained an analogous energy evolution for anharmonic cells with
energy-conserving noise. Bernardin et al.~\cite{BHLLO} wrote the formal
leading-order coefficient of noisy cells as the independent-cell term minus
a correction from the slow evolution of the cell energies. For
one-coordinate cells with pinning potential $|x|^r/r$, $r>2$, which includes
the quartic case, their Proposition~8.1 shows that this coefficient remains
bounded as the rate of velocity flips, random reversals of the momentum,
goes to zero. The argument uses the fact that
resonances between neighboring oscillators, that is, neighbors whose
frequencies are nearly commensurate, occur with small probability in the
Gibbs state. The full deterministic Green--Kubo coefficient at finite
coupling remains open. De Roeck and Huveneers~\cite{DRH,DRH19} bound
truncated Green--Kubo integrals for rotor and nonlinear Schr\"odinger
chains, and transport in the rotor chain with added noise. Their argument
shows that, to any order in the coupling, the current averages out except
on rare resonant configurations; they describe its extension to quartic
oscillators with harmonic coupling, our separable chain, as an open
question. For cells with integrable dynamics, these bounds rest on the
rarity of resonant configurations in the Gibbs state. They do not give the time
dependence of the running integral after its early plateau.

Resonance theory reduces two weakly coupled nonlinear oscillators with
nearly equal frequencies to a pendulum for their relative
phase~\cite{Chirikov,LL,Basko}. For a chain of quartic oscillators with
harmonic coupling, which our model contains, Mulansky et
al.~\cite{Mulansky} associate strong chaos with triple resonances of
neighboring oscillators. The response of an ensemble of pendulums is also
classical. In nonlinear Landau damping, particles trapped by a wave
oscillate at a bounce frequency, and the damping rate oscillates on the
corresponding period~\cite{Mazitov,ONeil,Palmadesso}; collisions that
interrupt the trapped motion bring the damping back toward its linear
rate~\cite{Kaganovich}. Here we map this trapping response to the running
integral of a resonant pair of quartic oscillators. For the quartic
potential, the period of small pendulum oscillations does not depend on the
pair energy~\cite{Chirikov}, so the whole thermal ensemble shares one time
scale. The new result is a parameter-free
prediction of the time dependence of the running integral of a resonant
pair after the plateau. The integral decreases, becomes negative as
exchanged energy returns, and reaches a first minimum whose time and depth
the theory fixes. Earlier resonance-based analyses bound transport
coefficients or locate the onset of chaos; the pair theory gives a
quantitative curve that simulations can test directly.

We test the prediction on isolated pairs and on chains, with two
separately written simulation codes. Isolated pairs approach the predicted
curve as the coupling decreases. In chains, the first minimum follows the
predicted time scale but is shallower. At much later times the running
integral keeps a small positive value, which is resolved for the rotated
chain and less certain for the separable chain. Chains of strongly chaotic
cells show no comparable decrease, and energy-conserving noise largely
removes the decrease when it acts more often than once per pendulum period.
Section~\ref{sec:model} defines the model and the observables,
Sec.~\ref{sec:theory} derives the pair response, Sec.~\ref{sec:tests}
compares it with pair and chain simulations, Sec.~\ref{sec:noise} adds
noise, and Sec.~\ref{sec:discussion} discusses the scope of the results.

\section{Model and observables}
\label{sec:model}

\subsection{Chain and cell dynamics}
\label{sec:chainmodel}

We study a ring of $N$ cells with periodic boundary conditions at
temperature $T=1$, in units in which the mass, the lattice spacing and
the Boltzmann constant are 1. Each cell $i$ has coordinates
$(x_i,y_i)$ and momenta $\bm p_i=(p_{x,i},p_{y,i})$:
\begin{equation}
\begin{split}
H&=\sum_i\Big[\frac{|\bm p_i|^2}{2}+V(x_i,y_i)\Big]
+\frac{\varepsilon}{2}\sum_i (x_{i+1}-x_i)^2,\\
V(x,y)&=\frac{x^4+y^4}{4}+\alpha\,x^2y^2 .
\end{split}
\label{eq:chain}
\end{equation}
The intercell coupling $\varepsilon\ll1$ acts only on the $x$ coordinates.
The parameter $\alpha$ couples the two coordinates within a cell and
selects the dynamics of an isolated, uncoupled cell. We compare four
values.

At $\alpha=0$ the cell potential separates into one quartic term for each
coordinate. The $y$ motion of each cell is then decoupled from $x$ and from
the other cells, its energy is conserved, and transport is carried by a
chain of quartic $x$ oscillators. We call it the separable chain. At
$\alpha=3/2$, the isolated cell separates in the rotated coordinates
$u,v=(x\pm y)/\sqrt2$, with potential $(u^4+v^4)/2$. We call $u$ and $v$
the rotated modes and this chain the rotated chain. The coupling acts on
$x$, which contains both modes, so each rotated mode exchanges energy with
both modes of each neighbor. Expanding Eq.~(\ref{eq:chain}) shows that the
coupling also adds an on-site term $\varepsilon x_i^2$ to each cell, which
couples the two modes within a cell at order $\varepsilon$. We call the separable and rotated cells integrable,
meaning that the isolated cell is integrable. At $\alpha=4$ and 8 the
isolated cell is strongly chaotic: the finite-time test in Sec.~S3 of the
supplementary material classifies almost all canonical orbits as chaotic.
Table~\ref{tab:symbols} lists the principal symbols.

\begin{table}[!t]
\caption{Principal symbols and terms.}
\label{tab:symbols}
\small
\setlength{\tabcolsep}{3pt}
\begin{tabular}{@{}>{\raggedright\arraybackslash}p{1.45cm}>{\raggedright\arraybackslash}p{5.0cm}l@{}}
\toprule
Symbol & Meaning & Defined \\
\midrule
$\varepsilon$ & Intercell coupling & Eq.~(\ref{eq:chain}) \\
$\alpha$ & Intracell coupling of $x$ and $y$; $\alpha=0$, $3/2$, 4 and 8 give separable, rotated and strongly chaotic cells & Sec.~\ref{sec:chainmodel} \\
$u,\,v$ & Rotated modes of an $\alpha=3/2$ cell, $(x\pm y)/\sqrt2$ & Sec.~\ref{sec:chainmodel} \\
$\kappa(t_c)$ & Running Green--Kubo integral of a chain with cutoff time $t_c$ & Eq.~(\ref{eq:gk}) \\
$\kappa_{\rm bare}$ & Bare coefficient, the independent-cell term & Eq.~(\ref{eq:bare}) \\
$\sigma_\omega$ & Canonical standard deviation of the cell frequency & Sec.~\ref{sec:observables} \\
$\kappa_{\rm pair}(t_c)$ & Running integral of an isolated pair & Eq.~(\ref{eq:kpair}) \\
$\kappa_L(t_c)$ & Late-window value, the mean of $\kappa$ over $[t_c/2,t_c]$ & Eq.~(\ref{eq:window}) \\
$E,\,I,\,\theta$ & Energy, action and angle of one $x$ oscillator & Sec.~\ref{sec:pendulum} \\
$h(I),\,\omega$ & Energy as a function of the action; frequency $\omega=dh/dI$ & Sec.~\ref{sec:pendulum} \\
$P,\,\varphi$ & Half the action difference and the relative phase of a pair & Eq.~(\ref{eq:pendulum}) \\
$\Omega_0,\,T_{\rm lib}$ & Frequency and period of small librations & Eqs.~(\ref{eq:pendulum}), (\ref{eq:tlib}) \\
$T_{\rm ref}$ & Reference period: $T_{\rm lib}$ for separable and $\sqrt2\,T_{\rm lib}$ for rotated cells & Sec.~\ref{sec:pendulum} \\
$F(s)$ & Pendulum response of the pair integral, with $s=\Omega_0t$ & Eq.~(\ref{eq:pairresponse}) \\
$t_{\min}$ & Time of the first minimum after the plateau & Sec.~\ref{sec:response} \\
$\gamma$ & Rate of energy-conserving momentum rotations & Sec.~\ref{sec:noise} \\
\bottomrule
\end{tabular}
\end{table}

Since only the $x$ coordinates couple neighboring cells, the total heat
current is
\begin{equation}
\begin{split}
J&=-\frac{\varepsilon}{2}\sum_i(p_{x,i}+p_{x,i+1})(x_{i+1}-x_i)\\
&=\sum_i\frac{\varepsilon}{2}\big(p_{x,i+1}x_i-p_{x,i}x_{i+1}\big).
\end{split}
\label{eq:current}
\end{equation}
The two forms differ by a sum that vanishes on a ring. Each term of either
sum is the energy current across one bond.

\subsection{Running integral and bare coefficient}
\label{sec:observables}

The running Green--Kubo integral with cutoff time $t_c$ is
\begin{equation}
\kappa(t_c)=\frac{1}{T^2N}\int_0^{t_c}C(s)\,ds,\qquad C(s)=\langle J(s)J(0)\rangle ,
\label{eq:gk}
\end{equation}
where $\langle\cdot\rangle$ is the canonical (Gibbs) equilibrium average at
temperature $T$. The conductivity is its limit $t_c\to\infty$, taken after
$N\to\infty$. At short times the running integral rises from zero and,
for weak coupling, approaches a plateau at finite time. We call it the
early plateau. Its value is close to the bare coefficient
\begin{equation}
\kappa_{\rm bare}=\frac{\varepsilon^2}{T^2}\int_0^\infty\langle p_x(t)x(0)\rangle^2\,dt ,
\label{eq:bare}
\end{equation}
where $\langle p_x(t)x(0)\rangle$ is the correlation of one isolated cell in
the canonical ensemble. Equation~(\ref{eq:bare}) is the
Green--Kubo integral of one bond current when the two neighboring cells
evolve independently, each with its uncoupled dynamics and its own
canonical distribution. We call it the independent-cell term. It measures
how fast the coupling transfers energy across a bond while the motion of
each cell is still correlated with its initial phase. For integrable
cells, this transfer accumulates almost entirely between neighbors with
nearly equal frequencies. Bernardin et al.~\cite{BHLLO} wrote the formal
leading-order coefficient of weakly coupled cells with energy-conserving
noise as this term minus a correction from the slow evolution of the cell
energies [their Eq.~(6.20)]. For integrable cells, the plateau forms once
this correlation has decayed by dephasing across the canonical spread of
cell frequencies, after a time of order $1/\sigma_\omega$, where
$\sigma_\omega$ is the standard deviation of the cell frequency. In the
pair theory, it lasts until resonant pairs return energy, on the much
longer time scale $T_{\rm lib}$ derived below.

We compute the bare coefficient separately from the chain: analytically for
integrable cells and, for chaotic or noisy cells, by multiplying
correlations measured in two independent ensembles of isolated cells. The
product avoids the positive bias from squaring a single noisy correlation
estimate. Sections~S1 and S3 of the supplementary material give the
derivation, the estimators and the values. Unless stated otherwise, results are normalized by this bare
coefficient; its uncertainty is common to all points that use the same
reference.

\subsection{Isolated pairs and late windows}
\label{sec:pairs}

To test the pair theory directly, we also simulate isolated pairs of
separable cells. Each pair carries the on-site terms that a cell has in
the chain, so its two $x$ oscillators have the Hamiltonian
\begin{equation}
H_2=\sum_{i=1,2}\Big(\frac{p_{x,i}^2}{2}+\frac{x_i^4}{4}\Big)
+\varepsilon\big(x_1^2+x_2^2-x_1x_2\big).
\label{eq:pairH}
\end{equation}
The running integral of the pair is
\begin{equation}
\begin{split}
\kappa_{\rm pair}(t_c)&=\frac{1}{T^2}\int_0^{t_c}\langle j(s)j(0)\rangle_2\,ds,\\
j&=\frac{\varepsilon}{2}\big(p_{x,2}x_1-p_{x,1}x_2\big),
\end{split}
\label{eq:kpair}
\end{equation}
where $\langle\cdot\rangle_2$ is the canonical average for $H_2$ and $j$ is
the one-bond form of Eq.~(\ref{eq:current}). Its early plateau is also the
bare coefficient to leading order in the coupling, so pair and chain
integrals divided by the bare coefficient can be compared directly.

At late times, the running integral of a chain oscillates about a slowly
varying value. We
measure that value by the late-window value
\begin{equation}
\kappa_L(t_c)=\frac{2}{t_c}\int_{t_c/2}^{t_c}\kappa(t)\,dt ,
\label{eq:window}
\end{equation}
the mean of the running integral over the window $[t_c/2,t_c]$. Its errors
come from resampling whole replicas. These are finite-window
measurements. A conductivity requires the corresponding long-time and
large-system limits.

\subsection{Simulations}
\label{sec:simulations}

We integrate with the velocity Verlet scheme, normally at time step
$dt=0.01$ after Langevin equilibration, and record the current every 0.1
time units. Two separately written simulation codes, implementations A
and B, produce separate trajectories, and one analysis code processes
both. Sections~S3 and S4 of the supplementary material give the run
settings, harmonic and bare-rate validation, uncertainty estimation and
time-step controls.

\paragraph*{Ethical approval} This study did not involve human participants or animals.

\section{Resonant-pair theory}
\label{sec:theory}

\begin{figure*}[!t]
\includegraphics[width=17cm]{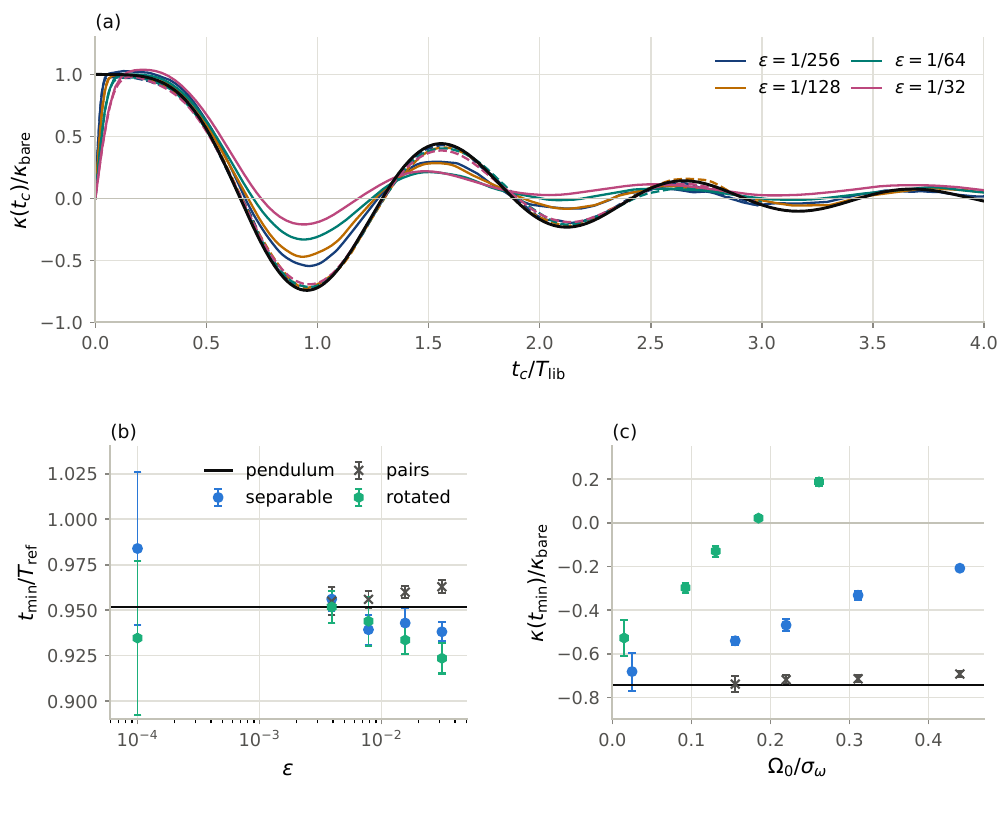}
\caption{Resonant-pair prediction and simulations. (a) Running integral
 divided by the bare coefficient against $t_c/T_{\rm lib}$ for separable
 chains (solid) and isolated pairs (dashed) at four couplings. Pair curves
 start from exact canonical samples of Eq.~(\ref{eq:pairH}); chain curves
 pool the available runs. The
 black curve is the pendulum response $F$ of Eq.~(\ref{eq:pairresponse}).
 (b) First-minimum time divided by the reference period, with
 $T_{\rm ref}=T_{\rm lib}$ for separable pairs and chains and
 $T_{\rm ref}=\sqrt2\,T_{\rm lib}$ for rotated chains; the horizontal line
 marks the prediction $\pendulumMinimumTime$. (c) First-minimum value
 against $\Omega_0/\sigma_\omega$, which equals $2.49\sqrt\varepsilon$ for
 separable and $2.49\sqrt\varepsilon/2^{3/4}$ for rotated cells. The
 horizontal line marks the prediction $\pendulumMinimumDepth$. Error bars
 are standard errors, by jackknife over groups of pairs and by bootstrap
 over chain replicas.}
\label{fig:integrable}
\end{figure*}

\subsection{Resonant neighbors set the bare coefficient}
\label{sec:golden}

For an integrable cell, the correlation in Eq.~(\ref{eq:bare}) is a
canonical average of oscillations at the cell frequency, which depends on
energy, and at its harmonics. Its square contains products of oscillations
at the frequencies of two independent cells, and the time integral of
such a product is proportional to a delta function of the difference
between the two frequencies. Only oscillations with equal frequencies
therefore contribute, and the bare coefficient is the golden-rule rate of
energy exchange between resonant neighbors. For the separable quartic
cell, 1:1 resonance of the fundamental frequencies accounts for all but
about $10^{-4}$ of the plateau.

\subsection{Pendulum normal form and libration period}
\label{sec:pendulum}

We derive the pair dynamics for the separable chain and then extend them
to the rotated chain. The parameter $\alpha$ enters through the way in
which an isolated cell separates into one-dimensional oscillators. At
$\alpha=0$, the $x$ coordinate of each cell is a quartic oscillator with
intercell coupling $\varepsilon$ and energy $E=p_x^2/2+x^4/4$. At
$\alpha=3/2$, each rotated mode is a quartic oscillator with intercell
coupling $\varepsilon/2$, as the end of this section shows. For $\alpha=4$ and 8, the
isolated cell does not separate and has no global action--angle variables,
so this reduction does not apply.

Write each $x$ oscillator in action--angle variables $(I,\theta)$, where
the action is the phase-space area of the orbit divided by $2\pi$. The
energy is a function $h(I)$ of the action, the frequency is
$\omega=dh/dI$, and the coordinate is a series of odd harmonics,
\[
x=A\sum_{n\ {\rm odd}}a_n\cos n\theta,\qquad A=(4E)^{1/4},
\]
where $A$ is the turning point and the fundamental coefficient is
$a_1=0.95501$. For neighbors $i$ and $i+1$ with nearly equal frequencies,
the coupling $-\varepsilon x_ix_{i+1}$ in Eq.~(\ref{eq:chain}) contains the slowly
varying term $-(\varepsilon A_iA_{i+1}a_1^2/2)\cos(\theta_i-\theta_{i+1})$.
The other terms oscillate at the cell frequency or faster, or describe
much weaker resonances between equal harmonics, and we drop them.
We average over the fast phase, expand $h(I_i)+h(I_{i+1})$ to second
order about the common action $I=(I_i+I_{i+1})/2$, and set
$A_iA_{i+1}\simeq A^2$, where $A$ is the turning point at energy $h(I)$.
This gives the approximate normal form
\begin{equation}
H_{\rm pair}=2h(I)+\omega'(I)P^2-\frac{\varepsilon A^2a_1^2}{2}\cos\varphi ,
\label{eq:pendulum}
\end{equation}
where $P=(I_i-I_{i+1})/2$ is conjugate to the relative phase
$\varphi=\theta_i-\theta_{i+1}$ and $\omega'=d\omega/dI=\omega\,d\omega/dE$.
The common action is conserved at this order, so Eq.~(\ref{eq:pendulum})
describes a pendulum for the relative phase~\cite{Chirikov,LL}; Chirikov
derived it for two coupled quartic oscillators~\cite[Sec.~3.3]{Chirikov}. A trapped
pair librates: its phase oscillates about zero, and energy flows back and
forth across the bond. An untrapped pair rotates: its phase runs through
all values, and its exchange averages out. Small librations have frequency
$\Omega_0=(\varepsilon\omega'A^2a_1^2)^{1/2}$.

For a quartic oscillator, $\omega=cE^{1/4}$ with $c=1.1981$ and
$A^2=2E^{1/2}$, so the product $\omega'A^2=c^2/2$ does not depend on
energy, and the period of small librations,
\begin{equation}
T_{\rm lib}=\frac{2\pi}{\Omega_0}=7.766\,\varepsilon^{-1/2},
\label{eq:tlib}
\end{equation}
is the same for all resonant pairs, whatever their energy. Chirikov's
frequency of small phase oscillations~\cite[Eq.~(3.45)]{Chirikov} has the
same energy independence; Eq.~(\ref{eq:tlib}) also keeps the coefficient
$a_1$ of the fundamental. We call
$T_{\rm lib}$ the libration period; the subscript stands for
libration. The on-site terms shift the frequency by a relative amount
of order $\varepsilon/\sqrt E$. We neglect them at leading
order; the expansion is nonuniform only at low energies, $E\lesssim\varepsilon^2$,
which carry little canonical weight.

The same derivation applies to each rotated mode, with one change that
sets its reference period. A rotated mode carries the
coupled coordinate only in part, $x=(u+v)/\sqrt2$. The bilinear coupling
between a mode of one cell and a mode of its neighbor, for example
$-(\varepsilon/2)u_iu_{i+1}$, therefore has half the strength of the
coupling between separable cells. The stiffer mode
potential $u^4/2$ raises $\omega'$ by a factor $\sqrt2$ and lowers $A^2$
by the same factor, so their product remains energy independent and
unchanged. The libration frequency $\Omega_0$ scales as the square root of
the coupling strength, so halving the coupling lowers it by $\sqrt2$ and
lengthens the reference period to
$T_{\rm ref}=\sqrt2\,T_{\rm lib}$. This reference leaves out the coupling
of the same order between the two modes within one cell.

\subsection{Parameter-free response of the pair integral}
\label{sec:response}

To leading order, the bond current of a resonant pair is proportional to
$\dot P$, the rate at which action passes from one oscillator to the
other. Its integrated autocorrelation is therefore proportional to
$\langle\dot P(0)[P(t)-P(0)]\rangle$. With the dimensionless time
$s=\Omega_0t$ and detuning $w=2\omega'P/\Omega_0$, which is the frequency
difference of the two oscillators in units of $\Omega_0$, the equations of
motion of Eq.~(\ref{eq:pendulum}) become
\begin{equation}
\frac{d\varphi}{ds}=w,\qquad \frac{dw}{ds}=-\sin\varphi .
\label{eq:dimless}
\end{equation}
These equations contain neither the energy nor the coupling. Since
$T_{\rm lib}$ does not depend on energy, every resonant pair in the thermal
ensemble follows this single dimensionless pendulum equation, without
parameters and on the same time scale, including pairs with large
librations and rotating pairs. The
thermal average therefore reduces to an average over the initial values
$(w_0,\varphi_0)$. For weak coupling, the resonance is narrow compared with
the spread of cell frequencies, so the canonical distribution of the
initial detuning is flat across it. Averaging over this
distribution gives
\begin{equation}
\begin{split}
\frac{\kappa_{\rm pair}(t_c)}{\kappa_{\rm bare}}&\simeq F(\Omega_0t_c),\\
F(s)&=\frac{2}{\pi}\int_{-\infty}^{\infty}dw_0
\int_{-\pi}^{\pi}\frac{d\varphi_0}{2\pi}\\
&\qquad\times\sin\varphi_0\,[w_0-w(s)] ,
\end{split}
\label{eq:pairresponse}
\end{equation}
where $w(s)$ solves Eq.~(\ref{eq:dimless}) from these initial values. The
integral runs over all initial detunings at fixed positive $s$; the
normalization $F(0^+)=1$ makes the pair integral start from the
bare coefficient. The same function describes the resonant particles,
trapped and passing, in a wave of fixed amplitude in nonlinear Landau
damping~\cite{ONeil,Palmadesso}, with the small-amplitude bounce frequency
replaced by $\Omega_0$. Section~S2 of the supplementary
material gives the numerical evaluation of this response and its
convergence checks.

The response stays close to 1 until about $0.2\,T_{\rm lib}$, which is
the early plateau at the left of Fig.~\ref{fig:integrable}(a). It then
falls, becomes negative and reaches its first minimum at
$t_{\min}=\pendulumMinimumTime\,T_{\rm lib}$ with depth
$\pendulumMinimumDepth\,\kappa_{\rm bare}$, followed by decaying
oscillations (black curve). The negative
values are coherent energy return: a librating pair sends energy back
across the bond, so the current at later times is anticorrelated with
the initial current. The running integral itself becomes negative: it is
proportional to the correlation between the initial current and the
action transferred since then, and once a librating pair has sent back
more action than it passed on, the transferred action has the sign
opposite to that of the initial current. A librating pair exchanges a bounded amount of
energy, so the long-time average of its running integral vanishes. The
theory fixes the time and the depth of the first minimum without fitted
parameters, in addition to the $\varepsilon^{-1/2}$ time scale.

\section{Comparison with simulations}
\label{sec:tests}

We compare the prediction first with isolated pairs and then with chains,
locating each first minimum by a local quadratic fit.

\subsection{Isolated pairs}
\label{sec:pairtests}

Simulated isolated pairs, Eqs.~(\ref{eq:pairH}) and (\ref{eq:kpair}),
include the on-site terms and the higher harmonics that the theory
neglects, so they measure finite-coupling corrections. Their running
integrals approach the pendulum response as the coupling decreases
[dashed curves in Fig.~\ref{fig:integrable}(a)]. At the weakest coupling,
$\varepsilon=1/256$, the first minimum is at $0.955\pm0.008\,T_{\rm lib}$
with depth $-0.739\pm0.036$ of the bare coefficient, in agreement with the
prediction at the measured precision. As the coupling grows, the minimum
becomes shallower and slightly later [Fig.~\ref{fig:integrable}(b) and
(c)]. We compare depths at each coupling
rather than pooling them into a single constant, since the finite-coupling
corrections differ between couplings.

\subsection{First minimum of chains}
\label{sec:chainminima}

The separable chain reaches its first minimum on the same time scale
[Fig.~\ref{fig:integrable}(b)]. From $\varepsilon=10^{-4}$ to 1/32, its
minimum time follows the predicted
$\varepsilon^{-1/2}$ scaling. At finite coupling, the chain prefactors
$t_{\min}/T_{\rm lib}$ also depart from the isolated-pair values. The
rotated chain follows the reference period
$T_{\rm ref}=\sqrt2\,T_{\rm lib}$, which the halved coupling between
rotated modes sets (Sec.~\ref{sec:pendulum}).

The chain minimum is shallower than that of an isolated pair at the same
coupling and becomes shallower as the coupling grows
[Fig.~\ref{fig:integrable}(c)]. The first-minimum value of the separable
chain rises from about $-0.7$ at $\varepsilon=10^{-4}$ to about $-0.2$ at
1/32. The horizontal coordinate
$\Omega_0/\sigma_\omega$ compares the width of the resonance with the
spread of cell frequencies, and the pair theory assumes that it is small.
At equal coupling it is smaller by $2^{3/4}$ for the rotated modes,
whose $\Omega_0$ is lower by $\sqrt2$ and whose frequency spread is larger
by $2^{1/4}$. These departures measure how far the pair
approximation holds in a chain. A cell in a chain can be close to resonance with
both of its neighbors at once. For the chain of $x$ oscillators of the
separable chain, Mulansky et al.~\cite{Mulansky} attribute strong chaos
mainly to such triple resonances, but the present minimum measurements do
not isolate their contribution. The rotated chain departs
further from the pair value and has an additional coupling between the
modes within each cell; its correction to the pair response remains
unresolved. Its first-minimum value is close to zero at
$\varepsilon=1/64$ and positive at 1/32.

\begin{figure*}[!t]
\includegraphics[width=17cm]{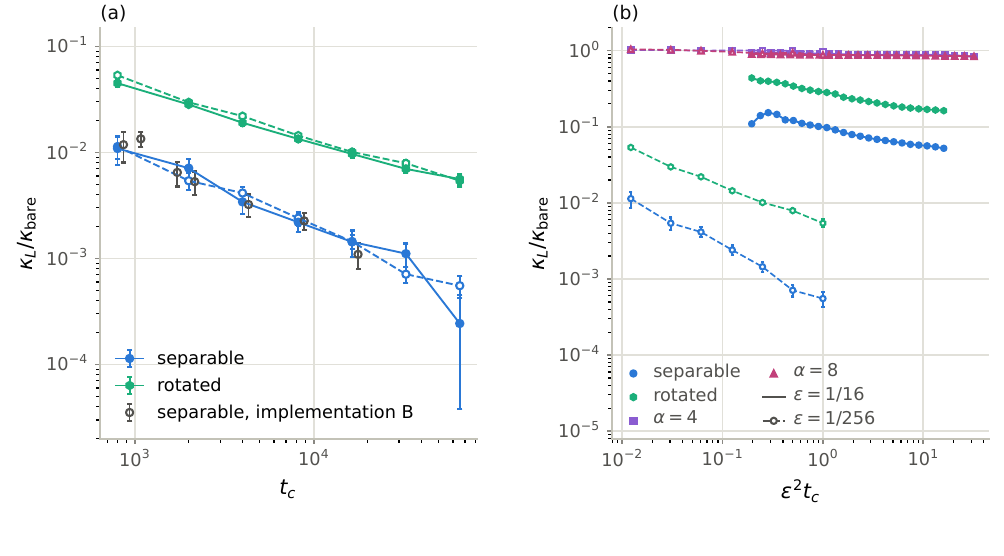}
\caption{Late-window values of chains. (a) Late-window value over
 $[t_c/2,t_c]$ at $\varepsilon=1/256$ for separable and rotated chains at
 $N=64$ (filled) and 128 (open), with a separate separable-chain run from
 implementation B (black open circles), displaced horizontally by a
 factor 1.08 for visibility. Lines connect measured values; they are solid
 for $N=64$ and dashed for 128. (b) The same quantity against
 $\varepsilon^2t_c$ for chains of integrable and strongly chaotic cells at
 $\varepsilon=1/16$ (solid) and 1/256 (dashed). At $\varepsilon=1/16$, the
 first windows include the time of the pair minimum. Error bars are
 standard errors and include the shared uncertainty of the bare
 coefficient where it is measured.}
\label{fig:retention}
\end{figure*}

\subsection{Late windows}
\label{sec:late}

The conductivity is the long-time limit of the running integral, so the
question after the first minimum is whether the running integral of a
chain returns to zero, as that of an isolated pair does, or keeps a
positive value. After the first minimum, it oscillates about a small value
that changes slowly with $t_c$. Late-window values, Eq.~(\ref{eq:window}),
follow that value at $\varepsilon=1/256$ up to $t_c=65536$
[Fig.~\ref{fig:retention}(a)]. Isolated pairs give a value consistent with
zero, as expected for librating pairs. In the final window, the rotated
chain keeps a small positive value, about $5\times10^{-3}$ of the bare
coefficient at both chain sizes and well above its errors. The separable
chain keeps about a tenth of that at $N=128$ and a smaller value at $N=64$
that is not resolved from zero, so the evidence for it is weaker. The
isolated-pair runs end at $t_c=16384$; in their last window the separable
chain from implementation B exceeds them by about 2.7 standard errors. The
late-window values decrease with $t_c$; Section~S3 gives their values and
their heavy-tailed replica distributions.

\subsection{Chains of strongly chaotic cells}
\label{sec:chaotic}

Chains with $\alpha=4$ and 8 show no comparable early decrease on the time
scale $T_{\rm lib}$. At $\varepsilon^2t_c=8$, they keep about 0.9 of their
independently measured bare coefficients at
$\varepsilon=1/16$ and 1/32 (the latter coupling is not shown).
Figure~\ref{fig:retention}(b) shows their late-window values with those of
the integrable-cell chains against $\varepsilon^2t_c$, that is, time in
units of $\varepsilon^{-2}$, the time scale of energy exchange between
neighboring cells. Within the errors, a slow further decrease remains
possible. This comparison shows an association between
strongly chaotic cells and the absence of coherent energy return; it
does not establish that orbital chaos is the cause. The chaotic cells differ from the integrable cells in three
ways at once. First, their orbits are chaotic. Second, the spectrum of $x(t)$ is
broadband rather than a set of discrete lines at the mode frequencies and
their harmonics. Third, an isolated chaotic cell conserves only its total
energy, whereas an isolated separable cell also conserves its $y$ energy
and an isolated rotated cell its two mode energies. The present comparison
changes all three together, so it cannot separate their contributions.

\begin{figure}[!t]
\includegraphics[width=8.5cm]{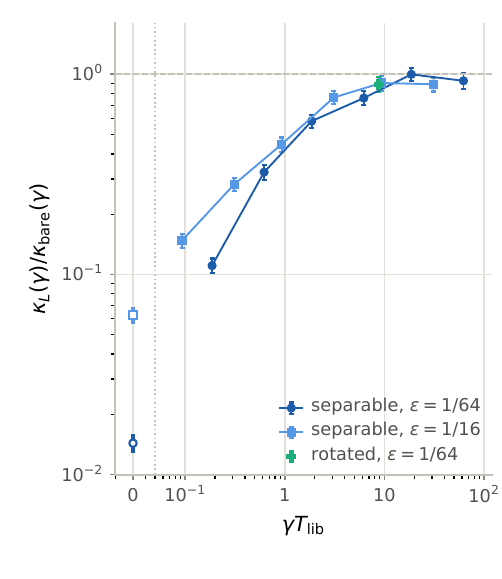}
\caption{Late-window value against the rate of energy-conserving noise.
 The late-window value over $t_c\in[1000,2000]$ is divided by the bare
 coefficient at the same noise rate, for the separable chain at
 $\varepsilon=1/64$ and 1/16 and the rotated chain at $\varepsilon=1/64$.
 The horizontal coordinate is $\gamma T_{\rm lib}$
 ($\gamma\sqrt2\,T_{\rm lib}$ for rotated cells). Open symbols at 0 are
 zero-noise values of the separable chain, displaced for the logarithmic
 axis; the dotted line separates them from the noisy measurements.
 The dashed horizontal line marks unity. Error bars combine the
 uncertainties of the late-window value and of the bare coefficient;
 points that share a bare coefficient share its uncertainty.}
\label{fig:noise}
\end{figure}

\section{Energy-conserving noise}
\label{sec:noise}

If the decrease after the plateau comes from coherent energy return,
perturbations that randomize the motion of a resonant pair faster than it
librates should remove it. We test this with energy-conserving momentum
rotations. At rate $\gamma$, each cell's momentum vector is rotated
through a uniformly random angle, which preserves
the cell energy and the canonical distribution. This is one of the energy-conserving
noises of Bernardin et al.~\cite{BHLLO}, which renews the direction of the
momentum; their bound for one-coordinate quartic cells uses velocity flips
instead. The noise interrupts a
libration after a time of order $1/\gamma$, so the plateau value should be
recovered when $\gamma T_{\rm lib}$ is of order one or larger. The rotations
also change the bare coefficient, so the chain is
normalized by the bare coefficient measured with the same noise.

The late-window value rises with the noise rate (Fig.~\ref{fig:noise}).
For the separable chain at $\varepsilon=1/64$, it rises from $0.11\pm0.01$
at $\gamma T_{\rm lib}=0.19$ to $1.00\pm0.07$ at 19. The second coupling
shows the same recovery trend, and the rotated chain, whose cells have no
uncoupled $y$ coordinate, is also close to its plateau at a high noise
rate. These finite-window observations are
consistent with noise interrupting coherent energy return on the time scale
$T_{\rm lib}$.

The scan locates the recovery scale only in order of magnitude.
Independent ensembles at the crossover show between-ensemble variation
beyond their internal errors (Sec.~S4), and the two couplings are observed
at different $\varepsilon^2t_c$. The precise recovery threshold and its
scaling remain unresolved.

\section{Discussion}
\label{sec:discussion}

In the weakly coupled quartic chain, the early plateau of the running
Green--Kubo integral is close to the bare coefficient, and the first
decrease after it occurs on the time scale on which resonant pairs return
exchanged energy. At weak coupling, the running integral of a resonant pair
follows the classical
pendulum trapping response, and the quartic potential gives this response
a single time scale, $T_{\rm lib}=7.766\,\varepsilon^{-1/2}$, for the whole
thermal ensemble. The theory predicts the time and depth of the first
minimum of an isolated pair without fitted parameters. Direct pair
simulations approach it as the coupling decreases and resolve
finite-coupling corrections. Chains follow the same time scale with
shallower minima, and at much later times they keep a small positive
value, resolved for the rotated chain and less certain for the separable
chain.

The late-window values, the chaotic-cell comparison and the noise scan
are finite-window measurements, and they leave the asymptotic conductivity
open. The chaotic-cell comparison associates strongly chaotic cells
with the absence of coherent energy return but does not identify orbital chaos as
its cause, since these cells also differ in their spectra and conserved
quantities. The present measurements do not
isolate the mechanism of the shallower chain minima, including the role of
triple resonances.

Quartic scaling connects these observations to high-temperature pinned
chains. Rescaling coordinates, momenta and time maps coupling
$\varepsilon$ at $T=1$ to coupling 1 at $T=\varepsilon^{-2}$; the chain
of $x$ oscillators of the separable chain is then the $\phi^4$ chain of
Aoki and Kusnezov~\cite{AK,AK02} at temperature $\varepsilon^{-2}$, above
the temperatures that they simulated. The libration period supplies
an observation scale for that chain. When
$\Omega_0/\sigma_\omega=2.49\sqrt\varepsilon\ll1$, a running integral
observed after the dephasing time $1/\sigma_\omega$ but well before
$T_{\rm lib}$ approaches the bare coefficient, and its subsequent decrease
reaches a first minimum near $\pendulumMinimumTime\,T_{\rm lib}$, the time
scale of coherent energy return in resonant pairs.

\section*{Supplementary Material}

See the supplementary material for the bare-coefficient derivation (S1),
the resonant-pair response and its numerical evaluation (S2), simulation
and statistical methods (S3), and numerical controls including noise
reproducibility (S4).

\section*{Acknowledgments}
J.~Olszewski acknowledges support from the University of Illinois Chicago
Chancellor's Undergraduate Research Award (CURA).

Claude Opus 5.5 (Anthropic) and Codex 6.1 (OpenAI) assisted with scientific
discussion, literature verification, simulation-code development, numerical
calculations and data analysis, and provided editorial assistance with the
manuscript under the authors' direction. Numerical outputs were checked
using harmonic and bare-rate benchmarks, comparisons between separate
simulation implementations, and time-step controls reported in Secs.~S3
and S4 of the supplementary material. The authors reviewed and edited all
AI-assisted material and take full responsibility for the final
interpretation and the content of this manuscript.

\section*{Conflict of interest}

The authors have no conflicts to disclose.

\section*{Author contributions}

\textbf{Julian Olszewski:} Conceptualization (equal); Funding
acquisition (lead); Investigation (equal); Methodology (equal); Writing
-- original draft (equal); Writing -- review \& editing (equal);
Software; Formal analysis; Data curation; Visualization.

\textbf{Zijian Liu:} Conceptualization (equal); Investigation (equal);
Methodology (equal); Writing -- original draft (equal); Writing --
review \& editing (equal); Software; Formal analysis; Data curation;
Visualization.

\textbf{Jie Xu:} Conceptualization (equal); Supervision (lead); Writing
-- review \& editing (equal).

\section*{Author responsibility statement}
All authors have accepted responsibility for the entire content of this manuscript and approved its submission.

\section*{Data availability}

The numerical source data are available in Mendeley Data~\cite{Dataset}.

\end{document}


\twocolumn[{
\begin{center}
{\Large\bfseries Supplementary Material: Resonant-pair energy return beyond the early Green--Kubo plateau in a quartic chain\par}
\vspace{10pt}
Julian Olszewski, Zijian Liu, Jie Xu\textsuperscript{*}\par
\vspace{5pt}
{\small Department of Mechanical and Industrial Engineering,\\
University of Illinois Chicago, 842 West Taylor Street (MC 251),\\
Chicago, Illinois 60607, USA\par}
\vspace{5pt}
{\small\textsuperscript{*}Corresponding author: \href{mailto:jiexu@uic.edu}{jiexu@uic.edu}\par}
{\small\textit{Author note.} Julian Olszewski and Zijian Liu contributed equally to this work.\par}
\end{center}
\vspace{14pt}
}]
\thispagestyle{plain}

\section{Bare coefficient of the quartic cell}
\label{sec:bare}

We evaluate explicitly, for the quartic cell without noise, the
independent-cell term of Bernardin et al.~\cite{BHLLO}, which their
Appendix A bounds for velocity-flip noise. The isolated cell with
$V=x^4/4$ has the exact solution
$x(t)=A\,\mathrm{cn}(At\,|\,\tfrac12)$ with energy $E=A^4/4$, since
$\mathrm{cn}''=-\mathrm{cn}^3$ at parameter $\tfrac12$. Its frequency is
$\omega=\pi A/[2K(\tfrac12)]=cE^{1/4}$ with $K(\tfrac12)=1.85407$ and
$c=\pi\sqrt2/[2K(\tfrac12)]=1.19814$. The Fourier series
\begin{equation}
\mathrm{cn}(u\,|\,\tfrac12)=\sum_{n\ {\rm odd}}a_n\cos\frac{n\pi u}{2K},\quad
a_n=\frac{2\pi\sqrt2}{K}\,\frac{e^{-n\pi/2}}{1+e^{-n\pi}},
\label{eq:cnseries}
\end{equation}
gives $a_1=0.95501$ and $a_3/a_1=0.045$. At temperature $T=1$ the cell energy
has the canonical density $\rho(E)=E^{-1/4}e^{-E}/\Gamma(3/4)$. Averaging over
the phase gives
$\langle p_x(t)x(0)\rangle=-\int_0^\infty S(\Omega)\sin\Omega t\,d\Omega$, where
$S$ collects the weight of each harmonic $n\omega(E)=\Omega$,
\begin{equation}
S(\Omega)=\sum_{n\ {\rm odd}}\rho(E_n)\frac{dE_n}{d\Omega}\,
\frac{A(E_n)^2a_n^2\,\Omega}{2},\qquad n\,\omega(E_n)=\Omega .
\label{eq:spectral}
\end{equation}
The bare coefficient $\kappa_{\rm bare}=\varepsilon^2\int_0^\infty\langle p_x(t)x(0)\rangle^2dt$ at $T=1$ then becomes
$\kappa_{\rm bare}=(\pi\varepsilon^2/2)\int_0^\infty S(\Omega)^2d\Omega$. With
the fundamental alone the integral is elementary,
\begin{equation}
\frac{\kappa_{\rm bare}}{\varepsilon^2}=\frac{\pi}{2}\,
\frac{4a_1^4c}{\Gamma(3/4)^2}\,\frac{\Gamma(11/4)}{2^{11/4}}=0.99700 ,
\label{eq:kbareclosed}
\end{equation}
and all harmonics together give 0.99712. The frequency spread
$\sigma_\omega=c\,[\Gamma(5/4)/\Gamma(3/4)-\Gamma(3/4)^{-2}]^{1/2}=0.32534$
follows from the same density. In the rotated cell $\alpha=3/2$,
$x=(u+v)/\sqrt2$ with $u$ and $v$ independent oscillators of potential $u^4/2$,
so $\langle p_x(t)x(0)\rangle$ equals the correlation of one such oscillator.
Its amplitude and frequency scale as $2^{-1/4}$ and $2^{1/4}$ relative to the
separable cell, which multiplies $\kappa_{\rm bare}$ by $2^{-3/4}$ and gives
0.5929. Its rotated-mode frequency spread is
$\sigma_{\omega,\mathrm{rot}}=2^{1/4}\sigma_\omega=0.3869$.

\section{Resonant-pair response}
\label{sec:pairensemble}

For the averaged pair Hamiltonian
$H_{\rm pair}=\omega'P^2-V_0\cos\varphi$, with
$V_0=\varepsilon A^2a_1^2/2$, define
$\Omega_0=(2\omega'V_0)^{1/2}$, $s=\Omega_0t$ and
$w=2\omega'P/\Omega_0$. The equations become
$d\varphi/ds=w$ and $dw/ds=-\sin\varphi$; the small-amplitude reference
period is $T_{\rm lib}=2\pi/\Omega_0$.
The current is proportional to $\dot P$, so its integrated correlation is
proportional to $\langle\dot P(0)[P(t)-P(0)]\rangle$.
The response $F(s)$ in the main text averages this expression over a locally
flat resonant measure $dw_0\,d\varphi_0/(2\pi)$. This measure is unnormalized
on the infinite momentum axis. Its normalization follows from the
large-detuning, freely rotating limit,
$F_{\rm rot}(s)=\pi^{-1}\int_{-\infty}^{\infty}\sin(w_0s)/w_0\,dw_0=1$
for $s>0$. The infinite-domain limit is taken at fixed positive $s$;
a finite-domain integral is zero at exactly zero time.

The correspondence can be seen directly from Liouville evolution.
At fixed common action, bias a stationary pair ensemble infinitesimally by
$\rho_\eta\propto\rho_{\rm eq}e^{\eta P}$, where $\eta$ is conjugate to
the action difference. Stationarity and
$\langle P\dot P\rangle_{\rm eq}=0$ give
\begin{equation}
\left.\frac{\partial\langle\dot P(t)\rangle_\eta}{\partial\eta}\right|_0
=\langle P(0)\dot P(t)\rangle_{\rm eq}
=-\langle\dot P(0)[P(t)-P(0)]\rangle_{\rm eq}.
\label{eq:responseidentity}
\end{equation}
For a current $J=a\dot P$, with $a$ constant on that shell, the expression
on the right is $-a^{-2}\int_0^t\langle J(0)J(u)\rangle_{\rm eq}du$.
In the weak-coupling resonant layer, the equilibrium weight is locally flat
in $P$ and the bias supplies a local linear slope. Evolving that slope with
the same pendulum flow is the prescribed-wave calculation used for
resonant particles~\cite{Mazitov,ONeil}. Rescaling to $w,s$ therefore gives
the same normalized kernel $F(s)$ and the same fixed-positive-time limit.
The heat-current coefficient and plasma damping rate retain their own
physical prefactors.

This is the fixed-amplitude pendulum trapping response of nonlinear Landau
damping~\cite{ONeil,Palmadesso}. The phase-averaged spectra below
reproduce O'Neil's Eq.~(31) for $\gamma(t)/\gamma_L$; his constant-field
approximation assumes $|\gamma_L|/\Omega_0\ll1$. Setting the integration
variable of Eq.~(55) in Palmadesso's report version~\cite{PalmadessoReport}
to $r=\sqrt m$ for trapped orbits and $r=2/v$ for passing orbits, they also
reproduce the time derivative of that equation.
The correspondence uses the common oscillatory integral, including its
fast-rotation tail, and describes a prescribed-amplitude response.

For explicit phase averaging, let $E_p=w^2/2+1-\cos\varphi$.
Librations have $E_p=2m$, $0<m<1$, and frequency
$\omega_L=\pi/[2K(m)]$; rotations have $v=(2E_p)^{1/2}>2$,
$m=4/v^2$ and $\omega_R=\pi v/[2K(m)]$.
Here $K$ is the complete elliptic integral of the first kind and
$q=\exp[-\pi K(1-m)/K(m)]$ is the elliptic nome.
With $K=K(m)$ and $q=q(m)$, the oscillatory momentum coefficients are
\begin{equation}
\begin{split}
b_{L,n}&=\frac{4\pi}{K}\frac{q^{n/2}}{1+q^n},\qquad n\ {\rm odd},\\
b_{R,n}&=\frac{2\pi v}{K}\frac{q^n}{1+q^{2n}},\qquad n\ge1 .
\end{split}
\label{eq:pendulumcoefficients}
\end{equation}
The constant part of the rotational momentum drops out of the
time derivative of its autocorrelation. That derivative, with the
Liouville measure, gives the equivalent quadrature
\begin{equation}
\begin{split}
F(s)&=\frac{2}{\pi}\left[\int_0^1 K_L(m,s)\,dm
+\int_2^\infty K_R(v,s)\,dv\right],\\
K_L&=\sum_{n\ {\rm odd}}n b_{L,n}^2\sin(n\omega_Ls),\\
K_R&=v\sum_{n\ge1}n b_{R,n}^2\sin(n\omega_Rs).
\end{split}
\label{eq:pendulumspectral}
\end{equation}
For fixed $s$, the fast-rotation kernel has the expansion
\begin{equation}
\begin{split}
K_R(v,s)\sim{}&
\frac{\sin(vs)}{v}-\frac{s\cos(vs)}{v^2}\\
&+\frac{(2-s^2/2)\sin(vs)}{v^3}\\
&+\frac{(s^3/6-13s/4)\cos(vs)}{v^4}+\cdots .
\end{split}
\label{eq:pendulumtail}
\end{equation}

We integrate the libration kernel by Gauss--Legendre quadrature and the
rotation kernel over $2<v<256$, retaining 128 harmonics in each branch
(odd harmonics for libration). The tail beyond 256 is restored with the
four terms of Eq.~(\ref{eq:pendulumtail}), with no fitted constants.
Cutoffs 64, 128 and 256, quadrature refinements and harmonic truncations give
$t_{\min}/T_{\rm lib}\simeq0.952$ and $F_{\min}\simeq-0.74$ at the
precision quoted in the main text.
An independent initial-condition quadrature with exact elliptic trajectories,
checked against a high-order ODE solver and step-refined leapfrog, verifies
the trajectory calculation. Convergence differences measure numerical
accuracy.

\section{Simulation and statistics}
\label{sec:methods}

\paragraph*{Sampling and runs.} Unless stated otherwise, $T=1$, velocity
Verlet uses $dt=0.01$ after 300 time units of Langevin equilibration with
friction 1, and current samples are separated by 0.1. The autocorrelation
uses a fast Fourier transform, divides each lag by its number of overlapping
samples and is integrated by the trapezoidal rule, without subtracting a
sample mean. Recording times are at least 3.75 times the maximum lag.
Table~\ref{tab:runs} summarizes run settings; the separate
noise-step controls in Table~\ref{tab:noisestep} complement the plotted results. Implementation B was written from the
stated methods without reusing A's simulation code; its developer had seen A.
The same analysis processes both outputs.

The isolated-cell classification uses 400 canonical orbits integrated to
$t=6000$ with two tangent vectors. An orbit is classified as chaotic when
its smaller alignment index, the smaller norm of the sum and difference of
two normalized tangent vectors, falls below $10^{-8}$ by that time; the
remaining orbits are classified as regular~\cite{Skokos01,SALI}.
For $\alpha=4$ and 8 the regular fractions are $0.003\pm0.002$ and
$0.005\pm0.004$, respectively. This is a finite-time classification.

\begin{table*}
\caption{Simulation run settings. $R$ counts independent replicas and $N$
cells per replica; bare-cell and canonical rows instead count groups
and cells per group. Pairs contain two cells each. Ranges summarize actual
runs.
Canonical rows use independent Verlet and correlation code. The paired
noisy row includes controls and four additional 64-replica blocks at each
step, using a separate noise driver.}
\label{tab:runs}
\begin{tabular}{llccc}
\toprule
cells and implementation & $\varepsilon$ & $R\times N$ & $t_{\rm rec}$ & $t_{\max}$ \\
\midrule
quartic size scan (A, $\alpha=0$) & $1/256$ to $1/4$ & $256\times64$, $128\times128$, $64\times256$ & 3000 & 800 \\
quartic chains (A, $\alpha=0,3/2$) & $1/256$, $1/64$ & $128\times64$ & 16000 & 4000 \\
quartic chains (A, $\alpha=4,8$) & $1/256$ to $1/4$ & $128\times64$ & 16000 & 4000 \\
extended chains (A, $\alpha=0,3/2,4,8$) & $1/256$ & $128\times64$, $128\times128$ & 262144 & 65536 \\
weak chains (B, $\alpha=0,3/2,4,8$) & $10^{-4}$ & $128\times64$ & 8000, 12000 & 1000, 3000 \\
quartic chains (B, $\alpha=0,3/2$) & $1/128$ to $1/16$ & $128\times64$, $256\times64$ & 8000, 32768 & 2000, 4096 \\
long chains (B, $\alpha=0,3/2$) & $1/256$ & $128\times64$ & 16000, 65536 & 4000, 16384 \\
quartic chains (B, $\alpha=4$) & $1/32$ to $1/8$ & $256\times64$ & 32768 to 131072 & 2048 to 16384 \\
quartic chains (B, $\alpha=8$) & $1/32$, $1/16$ & $256\times64$ & 16384, 65536 & 2048, 8192 \\
equilibrated pairs (B) & $1/256$ to $1/32$ & $128\times128$ & 8000 to 65536 & 2000 to 16384 \\
noisy scan (A, $\alpha=0,3/2$) & $1/64$, $1/16$ & $128\times64$ & 8000 & 2000 \\
harmonic chains (A) & $1/2$, $1$ & $128\times128$ to $16\times1024$ & 3000 & 800 \\
noisy bare cells (independent) & 0 & $16\times128$ & 8192 & 1024 \\
canonical cells ($\alpha=4,8$; independent) & 0 & $48\times512$, $16\times512$ & 4096 & 200 \\
canonical pairs (independent) & $1/256$ to $1/32$ & $16\times4096$ & $12T_{\rm lib}$ & $3T_{\rm lib}$ \\
paired noisy chains (independent) & $1/64$ & $256\times64$ & 8000 & 2000 \\
\bottomrule
\end{tabular}
\end{table*}

\paragraph*{Bare coefficient and error propagation.} Deterministic
references use exact canonical samples at $T=1$: 24576 cells for $\alpha=4$
and 8192 for $\alpha=8$, split into groups of 512. The homogeneous polar
sampler draws angle with density proportional to the inverse square root
of the angular quartic potential, radial potential energy from a
$\Gamma(1/2,1)$ law, and Gaussian momenta. Velocity Verlet with $dt=0.005$
records 4096 time units. Products of different groups estimate
$\langle p_x(t)x(0)\rangle^2$ without the positive sampling-variance bias
from squaring one noisy estimate. The bare coefficient is the mean running
integral over $[50,100]$, with a leave-one-group-out jackknife error for
that same window. This gives $0.330\pm0.002$ and $0.226\pm0.003$ in units of
$\varepsilon^2$. Paired $dt=0.01$ and 0.005 runs use the same initial
states; their coefficient differences are smaller than their paired errors.
Step-refined copies are counted once in the reference sample.

Noisy references use 2048 cells recorded for 8192 time units, split into 16
groups. Their center uses the same $[50,100]$ window and their uncertainty
uses the final-lag jackknife at 1024. For the noisy separable cell the
coefficient decreases from $0.987\pm0.028$ at $\gamma=0.003$ to
$0.273\pm0.009$ at $\gamma=1$; the noisy rotated cell at $\gamma=0.1$
gives $0.504\pm0.015$.

For a late-window mean $L$ divided by an independent bare estimate $B$,
linear propagation gives
$\operatorname{var}(L/B)=\operatorname{var}(L)/B^2+
L^2\operatorname{var}(B)/B^4$.
A common bare estimate correlates ratios across couplings and cutoffs.
Any intermediate early plateau $E$ cancels in $(L/E)/(B/E)=L/B$;
its uncertainty cancels with the associated covariance.
Ratios to a run's own early plateau instead use a delete-one-replica
jackknife of $L/E$. Each replica's window mean is taken before resampling,
retaining correlations between lags. The early plateau $\kappa_2$ is the mean running chain integral over
$[3\tau,6\tau]$, with $\tau$ the first time $C(t)/C(0)$ falls below $1/e$.
Minimum-time and minimum-value estimates use a quadratic fit to the ensemble
curve over $\pm0.1$ pair periods around its lowest point, with whole-replica
bootstrap errors. Inverse-variance pooling of $n$ runs uses error scale
factor $\max[1,\sqrt{\chi^2/(n-1)}]$ where scatter requires it.
Quadratic windows of $\pm0.15$ and $\pm0.2$ periods and a cubic-spline
minimum support the half-power time scale.

\paragraph*{Direct canonical pairs.} At each of the four couplings,
32768 independent pairs (65536 cells) are grouped into 16 statistical groups. Rejection
sampling includes the full potential
$x_1^4/4+x_2^4/4+\varepsilon(x_1^2+x_2^2-x_1x_2)$, followed by independent
Gaussian momenta. Pair running integrals use the current autocorrelation
per pair, with the same per-bond bare reference as the chain. Curves use
$dt=0.01$, recording time $12T_{\rm lib}$
and lags through $3T_{\rm lib}$. The same quadratic minimum locator is
applied to each delete-one-group curve. Separate depths at $\varepsilon$
equal to 1/32, 1/64, 1/128 and 1/256 are
$-0.693\pm0.014$, $-0.715\pm0.016$, $-0.719\pm0.020$ and $-0.739\pm0.036$. Endpoint step-halving checks change the
depth by less than $5\times10^{-5}$ and the scaled minimum time by less
than $3\times10^{-5}$. Long-window pair residuals use the separate runs
listed in Table~\ref{tab:runs}.

\begin{figure*}
\includegraphics[width=17cm]{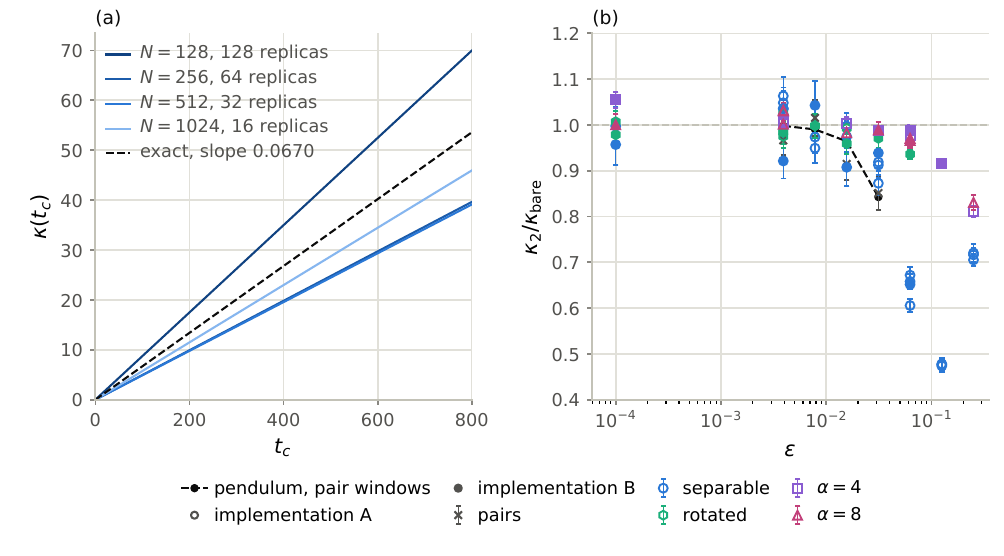}
\caption{Current and bare-rate validation. (a) Running Green--Kubo integral
of the pinned harmonic chain at $\varepsilon=1/2$ for four sizes, against
the exact slope of Eq.~(\ref{eq:harmonic}). The current is conserved, so
each size has the sampling error $(2/R)^{1/2}$ of its $R$ replicas.
(b) Early plateau $\kappa_2$ divided by the bare coefficient for the two
integrable chains, strongly chaotic cells and equilibrated isolated-pair
validation runs (crosses).
Open symbols denote implementation A and filled symbols B. Errors use
replica window means over the fixed estimated $[3\tau,6\tau]$ window and
independent bare-coefficient errors. The line averages the pendulum response
over those validation runs' measured windows.}
\label{fig:valid}
\end{figure*}

\paragraph*{Validation and independent implementations.}
Figure~\ref{fig:valid} tests the current estimator and compares the early
plateau with the independently computed bare coefficient. For a pinned
harmonic test chain the current is conserved and the exact slope is
\begin{equation}
\frac{C(0)}{NT^2}=\frac{1+2\varepsilon-\sqrt{1+4\varepsilon}}{4}.
\label{eq:harmonic}
\end{equation}
Each replica supplies one sample of $J^2$, so the slope has relative
sampling error $(2/R)^{1/2}$. At $\varepsilon=1/2$, pooling 240 replicas gives
$0.070\pm0.006$, against 0.0670; at $\varepsilon=1$ the result is
$0.182\pm0.017$, against 0.1910. For the separable chain,
$\kappa_2/\kappa_{\rm bare}$ ranges from 0.91 to 1.06 in 14 runs at
$\varepsilon\le1/64$. Larger couplings place part of the early window
inside the pair response (Sec.~S2).
Figure~\ref{fig:valid}(b) shows the chaotic-cell early plateaus with
their independent bare references and shared-denominator uncertainties.

The two implementations agree for some quantities and couplings, while
other measurements show additional run-to-run scatter. At
$\varepsilon=1/256$, the five separable-chain minimum depths are consistent
($\chi^2=3.1$ for four degrees of freedom). At $\varepsilon=1/64$, A gives
$-0.334\pm0.027$, $-0.342\pm0.020$ and $-0.379\pm0.029$, against
$-0.273\pm0.026$ from B; this scatter exceeds the within-run errors.
For the chaotic cell $\alpha=4$ at $\varepsilon=1/16$, the late ratio to the
run's own early plateau over $[2000,4000]$ is $0.947\pm0.069$ in A and
$0.890\pm0.022$ in B. Pooled estimates include a between-run scale factor
where required.

\begin{table*}[!t]
\caption{Paired noise-step controls. Entries give means and conditional
standard errors for the chain window mean $\kappa_L$, bare coefficient and
$r=\kappa_L/\kappa_{\rm bare}$. The last column is $10^3d_{\max}$, with
$d_{\max}=\max_r|H_r(t_{\rm rec})-H_r(0)|/H_r(0)$ across the 256
chain replicas. Each block couples its two steps; Tests 1 and 3 are
independent ensembles, and all three tests are separate from the
128-replica noise scan in Fig.~3 of the main text.}
\label{tab:noisestep}
\begin{tabular}{ccccccc}
\toprule
$\alpha$ & $\gamma$ & $dt$ & $\kappa_L/\varepsilon^2$ & $\kappa_{\rm bare}/\varepsilon^2$ & $r$ & $10^3d_{\max}$ \\
\midrule
\multicolumn{7}{l}{Test 1: separable cell} \\
0 & 0.03 & 0.01 & $0.528\pm0.029$ & $0.951\pm0.038$ & $0.555\pm0.038$ & 0.104 \\
0 & 0.03 & 0.005 & $0.433\pm0.025$ & $0.946\pm0.036$ & $0.457\pm0.032$ & 0.033 \\
\midrule
\multicolumn{7}{l}{Test 2: $\alpha=4$} \\
4 & 1 & 0.01 & $0.1365\pm0.0081$ & $0.1304\pm0.0037$ & $1.047\pm0.069$ & 2.864 \\
4 & 1 & 0.005 & $0.1386\pm0.0073$ & $0.1316\pm0.0037$ & $1.054\pm0.063$ & 0.537 \\
\midrule
\multicolumn{7}{l}{Test 3: independent separable ensemble} \\
0 & 0.03 & 0.005 & $0.554\pm0.031$ & $0.916\pm0.021$ & $0.605\pm0.037$ & 0.038 \\
0 & 0.03 & 0.0025 & $0.585\pm0.029$ & $0.912\pm0.021$ & $0.642\pm0.035$ & 0.007 \\
\bottomrule
\end{tabular}
\end{table*}

\begin{figure*}
\includegraphics[width=17cm]{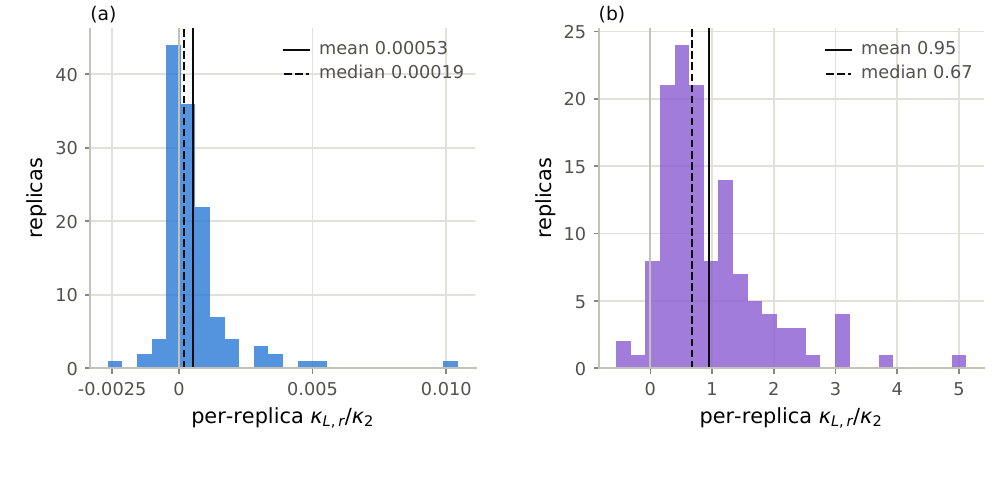}
\caption{Replica distributions of late-window values. Ratios
$\kappa_{L,r}/\kappa_2$ use each replica's window mean over $[32768,65536]$
and the common ensemble early plateau, at $\varepsilon=1/256$ and $N=128$.
(a) Separable chain. (b) Chaotic cell $\alpha=4$. Solid lines mark the mean;
dashed lines mark the median.}
\label{fig:replicas}
\end{figure*}

\paragraph*{Long-window statistics.} Divided by the bare coefficient, the
final-window values at $\varepsilon=1/256$ are $(5.4\pm0.6)\times10^{-3}$
at $N=128$ and $(5.6\pm0.7)\times10^{-3}$ at $N=64$ for the rotated chain,
and $(5.5\pm1.3)\times10^{-4}$ and $(2.4\pm2.1)\times10^{-4}$ for the
separable chain. Isolated pairs, recorded to $t_c=16384$, give
$(0.8\pm2.3)\times10^{-4}$ over $[8192,16384]$, where the separable chain
of implementation B at $N=64$ exceeds them by $(1.0\pm0.4)\times10^{-3}$,
or 2.7 standard errors. Figure~\ref{fig:replicas} shows the
broad replica distributions. At $\varepsilon=1/256$ in the final window
$[32768,65536]$, the five largest separable-chain replicas carry 41 percent
of the sum at $N=128$ and 76 percent at $N=64$; 41 of the 128
$N=128$ replicas have negative window means. Means therefore remain sensitive to rare contributions.
Whole-replica bootstrap errors agree with the jackknife errors for these
window means.
At $N=64$, whole-replica resampling gives the endpoint
$\kappa(65536)/\kappa_2=-0.021\pm0.034$; first averaging each replica over
$[32768,65536]$ gives $0.00024\pm0.00020$ with the same $\kappa_2$
normalization. Late-window values use this window estimator.

\section{Numerical controls and noise reproducibility}
\label{sec:noisestep}

For deterministic chain runs the maximum endpoint relative energy error
across replicas is $1.3\times10^{-4}$; it reaches $1.7\times10^{-3}$ in the
noise-rate scan at $\gamma=1$. In the deterministic $\alpha=8$ check at
$\varepsilon=1/16$, halving $dt$ to 0.005 lowers the energy error by a factor
of 3.5 and changes $\kappa_2/\varepsilon^2$ from 0.2191 to 0.2197.
The paired change of the late ratio to its own early plateau at $t_c=2048$
is $-0.031\pm0.051$. These runs share the equilibration random seed;
different Langevin step sizes give different recorded initial states.

Table~\ref{tab:noisestep} gives all six noise-step controls. Each uses
$N=64$, $T=1$, $\varepsilon=1/64$ and 256 independent replicas recorded
for 8000 time units, sampled every 0.1. Chain window means use
$[1000,2000]$. The independent matching bare ensemble has 16 groups of
128 cells recorded for 8192 time units, with lags to 1024 and a mean over
$[50,100]$. Here its uncertainty and cross-step covariance come from a
jackknife of that window mean, rather than the final-lag convention of
Sec.~S3. Ratio errors combine chain and independent bare uncertainties;
paired differences retain covariance across steps.

The two steps in each test share direct canonical initial states.
Coordinates drawn from $\exp[-(x^4+y^4)/4]$ are accepted with weight
$\exp[-\alpha x^2y^2]$, then complete chains with weight
$\exp[-(\varepsilon/2)\sum_i(x_{i+1}-x_i)^2]$; momenta are Gaussian.
A uniform-angle momentum rotation follows each full Verlet step with
probability $\gamma\,dt$. Coupled fine and coarse events preserve these
marginal probabilities: if the fine probability is $p$, the union of two
fine events has probability $2p-p^2$; a coarse event is added with
probability $p^2/(1-p)^2$ when neither occurs. The coarse angle is the sum
of fine angles, or an independent uniform angle for the added event.

For the separable chain at $\gamma=0.03$, Test 1 gives a paired change of $r$
(fine minus coarse) of $-0.098\pm0.043$, while Test 3 gives
$0.037\pm0.046$. The two independent $dt=0.005$ estimates differ by
$0.148\pm0.049$; their chain numerators differ by
$0.121\pm0.040$ in units of $\varepsilon^2$, both 3.0 combined standard
errors. The estimates of $r$ span 0.457 to 0.605, compared with
within-ensemble errors of 0.032 and 0.037. These ensembles reveal local
reproducibility variation beyond internal resampling errors. Step effects
and between-ensemble variation remain unseparated, so the quantitative
half-recovery threshold is unresolved. Each estimate and its conditional
error is retained separately.

Regenerating the first 16 replicas of the lower $dt=0.005$ ensemble
(Test 1) from the same seeds reproduces their initial states and currents
exactly, and an independent extraction of all recorded windows reproduces
the estimates. Four further blocks of 64 paired replicas with fixed seeds
give chain numerators $0.531\pm0.026$ at $dt=0.01$ and $0.519\pm0.029$ at
0.005, in units of $\varepsilon^2$, with paired difference
$-0.012\pm0.038$; the four-block 95-percent $t$ interval for the
difference is $[-0.109,0.086]$, so these blocks resolve no systematic step
shift. Combining the two independent $dt=0.005$ ensembles with the four
blocks at the same step gives a between-ensemble heterogeneity statistic
$Q=14.32$ on five degrees of freedom ($p=0.014$). The dispersion between
ensembles therefore remains; no ensemble is excluded, and these controls
provide no global error multiplier for the $dt=0.01$ scan.

For $\alpha=4$ at $\gamma=1$, Test 2 gives a paired change of $r$ of
$0.007\pm0.086$, with both estimates consistent with unity at this precision.
Its bare coefficient changes by
$(0.00115\pm0.00018)\varepsilon^2$, a resolved denominator dependence
included in the ratios. The maximum individual bare-cell endpoint energy
error falls from $1.47\times10^{-2}$ to $3.14\times10^{-3}$ on step halving.
Momentum rotations conserve physical energy; the monitored integration
error enters through composition with the Verlet update.
\begin{minipage}{\columnwidth}

\end{minipage}